\documentclass[slac_one]{revtex4}
\usepackage{graphicx}
\usepackage{fancyhdr}
\usepackage{epsfig}
\newcommand{\spar}{{\ensuremath\stackrel{\rightarrow}{\Rightarrow}}}
\newcommand{\sant}{{\ensuremath\stackrel{\rightarrow}{\Leftarrow}}}
\newcommand{\de}{{\rm\,d}}
\newcommand{\sparq}{{\ensuremath\stackrel{\rightharpoonup}{\Rightarrow}}}
\newcommand{\santq}{{\ensuremath\stackrel{\rightharpoonup}{\Leftarrow}}}
\newcommand{\detwo}{{\rm\,d^2\!}}

\newcommand{\Ftwo}   {\mbox{$\tilde{F}_2$}}

\newcommand{\Fz}   {\mbox{$\tilde{F}_3$}}

\newcommand{\Fem}  {\mbox{$F_2$}}

\newcommand{\Fint} {\mbox{$F_2^{\gamma Z}$}}
\newcommand{\Fwk}  {\mbox{$F_2^{Z}$}}

\newcommand{\Fzint} {\mbox{$F_3^{\gamma Z}$}}
\newcommand{\Fzwk}  {\mbox{$F_3^{Z}$}}

\newcommand{\QQ}  {\mbox{${Q^2}$}}

\newcommand{\bq}{\overline{q}}

\begin{document}

\title{Structure functions} 

%

\author{C. Diaconu}
\affiliation{Centre de Physique des Particules de Marseille and Deutsches Elektronen Synchrotron Hamburg}

\begin{abstract}
Recent progress in the understanding of the nucleon is presented. The unpolarised structure functions are obtained with unprecedented precision from the combined H1 and ZEUS data and are used to extract proton parton distribution functions via  NLO QCD fits. The obtained parametrisation displays an improved precision, in particular at low Bjorken $x$, and leads to precise predictions of cross sections for LHC phenomena.  Recent data from proton-antiproton collisions at Tevatron  indicate further precise constraints at large Bjorken $x$. The flavour content of the proton is further studied using final states with charm and beauty in DIS $ep$ and $p\bar{p}$ collisions. Data from polarised DIS or proton-proton collisions are used to test the spin structure of the proton and to constrain the polarised parton distributions.

\end{abstract}

\maketitle

\thispagestyle{fancy}


\section{Introduction} 
\begin{figure*}[t]
\centering
\includegraphics[width=95mm]{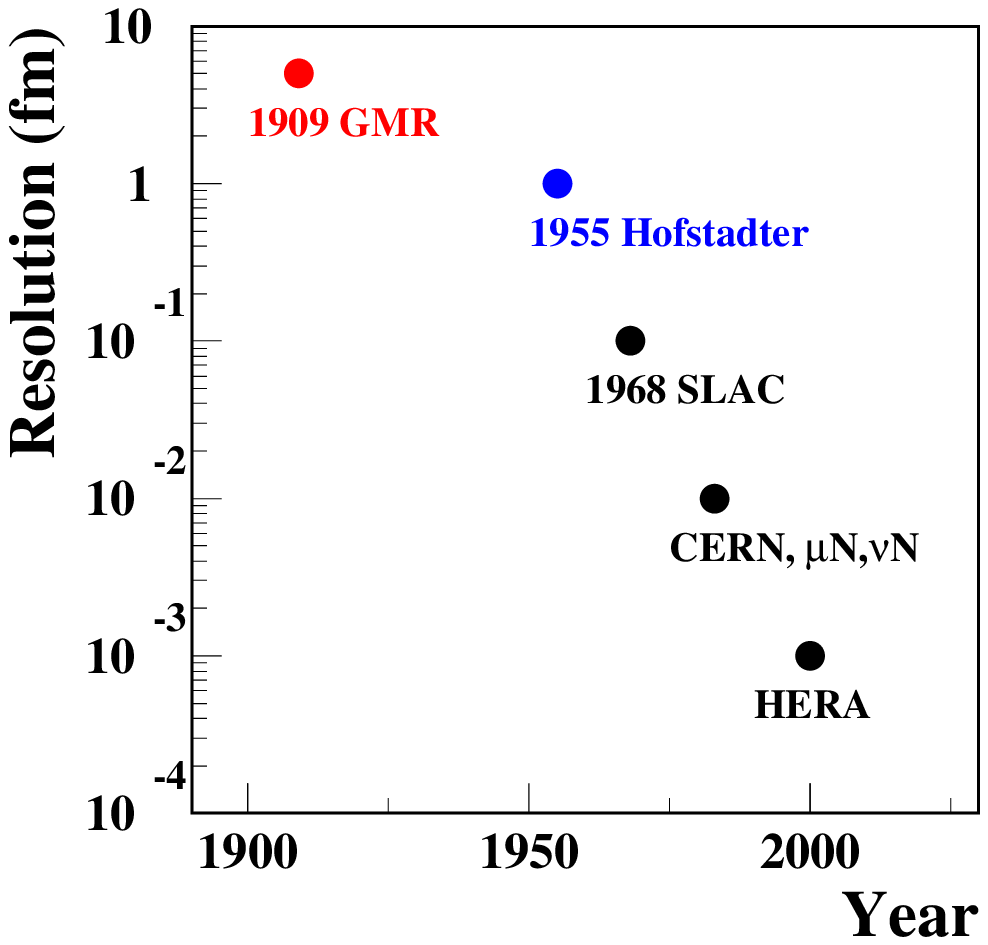}
\includegraphics[width=75mm]{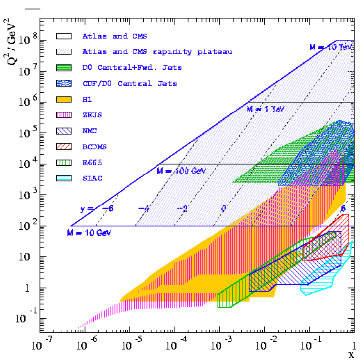}
\caption{Left: The historical developments of the matter investigations. Right: The kinematic plane $x$-$Q^2$, with the regions covered by the fixed target experiments, and the HERA and Tevatron colliders. The domain probed at LHC is also indicated.} \label{history}
\end{figure*}

The progress in understanding the structure of baryonic matter has been driven by the continuous improvement in the experimental ability to probe the matter with an increasing precision. Particle collisions in which  one of the particles  is preserved after the interaction is one of the most precise methods. For each collision the quantity of momentum transfered  from the probe to the composed particle can be measured. This momentum, denoted by $Q$, reflects the virtuality of the exchanged boson and corresponds to a wavelegth given by $\lambda {\mathrm{[fm]}}=
 0.2 / Q[\mathrm{MeV}] $. It gives the order of magnitude of the resolution with which the composed particle (e.g. the nucleon) is scrutinised in the collision - the higher the $Q$ the finer the details probed in the interaction.  A representation of the historical developments associated with the search for matter structure, from the discovery of the atomic nucleus in the classical Geiger-Marsden-Rutherford experiment, to the unique electron--proton collider HERA, is shown in figure~\ref{history} (left).
 For a momentum transfer of around 1~GeV the probed distances are comparable with the size of the proton. The inner structure of the proton was revealed in the first deep-inelastic scattering experiments at SLAC, 40 years ago. The ``partons" were found and later identified with quarks and gluons within the framework of quantum chromodynamics (QCD). The ``state of the art" is defined by the unique electron-proton collider HERA, where the centre-of-mass energy of 319~GeV allows the investigation of the proton with resolutions corresponding to $10^{-18}$~m.
 \par
 The status of the proton structure investigations can also be represented in the plane spaned by $Q^2$ and the variable $x$, representing the fraction of the proton momentum carried by the interacting parton, as in figure~\ref{history} (right). The unique reach of HERA  allows for the first time the investigation in the perturbative domain of the proton structure at very low $x$, down to $10^{-5}$. This regime is particularly important for LHC physics, since the production of particles with moderate masses (a few hundred GeV's) may involve partons at low $x$. The measurements are complemented by the proton-antiproton collider Tevatron, where the investigations of W boson production constrain the proton structure at high $x$.
Finally, the picture of the structure of the proton is also revealed in spin studies, performed using  collisions of polarised beams. These studies aim to measure the various components contributing to the proton's spin and also provide independent tests of the strong interaction mechanisms.
 
 \section{Proton structure and partons distribution functions}
 The proton can be described  in the infinite momentum frame as a collection of charged
 constituents (partons) that are "frozen" during the interaction time. In other words, in lepton-proton collisions, the time scale of the interaction between the incoming lepton and the proton's charged constituents is much shorter than the typical re-arrangement time of partons within the proton. Each interaction is characterised by the negative momentum transferred squared $Q^2$ and by the Bjorken $x$ variable, defined as the fraction of the proton momentum carried by the interacting parton.\par
The understanding of high energy collisions involving protons within the  QCD--improved quark parton model relies on the factorisation theorem, which states that the cross section is a convolution of a perturbative part, which describes the lepton-parton interaction at high energy, with a nonperturbative part that describes the momentum distributions of the proton inside the proton~\cite{Collins:1989gx}. In the DIS case:
\begin{equation} \label{xsec_dis}
\sigma_{DIS}(x,Q^2)= {\cal C}_{DIS} \otimes  f(x,Q^2)
\end{equation}
The coeficient functions ${\cal C}_{DIS}$ are calculable at any order in QCD, while the parton distribution functions $f(x,Q^2)$ (PDFs) encapsulate the parton content of the proton and are non-perturbative by definition. Their form cannot be calculated within QCD, nevertheless the evolution in $Q^2$ (i.e. their derivatives as a function of $Q^2$) are predicted by QCD via the DGLAP equations~\cite{dglap}. If the PDFs are parametrised as a function of $x$ at a given $Q^2=Q^2_0$, the evolution equations allow a recalculation at any $Q^2$ and therefore a prediction for any $(x,Q^2)$ point of the DIS cross section that depends only on the PDF $x$-parameterisation at $Q^2_0$. 
\par
The factorisation ansatz is usually also applied, though not completely demonstrated, for proton-proton or proton-antiproton collisions. In this case, the perturbative coefficient functions, describing the partonic interactions (between quarks and gluons) at high energies, are convoluted twice with the PDFs, corresponding to the two interacting (anti)protons. The interaction scale (the equivalent of $Q^2$ in DIS) can be given by the measurement of the final state. For instance, the cross section for $W$ boson production in $p\bar{p}$ collisions at Tevatron is expressed as:
\begin{equation} \label{xsec_wtev}
\sigma_{p\bar{p}\rightarrow W X }= \int dx_1 dx_2 {\cal C}_{\mathrm partons\rightarrow W } \otimes  f(x_1,M_W^2) \otimes f(x_2,M_W^2)
\end{equation}
The level of experimental precision achieved, together with a precise calculation of the perturbative coefficients allow these measurements to further constrain the PDFs in the domain at large $x$, as  will be described below.
\par
Measurements in a large domain in $x$ and $Q^2$, as indicated in figure~\ref{history}(right), from DIS $ep$ in fixed target and collisions at HERA, and from $p\bar{p}$ collisions at Tevatron are used to constrain the PDF parametrisation in a global fit.  Each measurement at a given point ($x$,$Q^2$) is compared with the prediction obtained from a parameterisation of the PDFs at a given scale $Q^2_0$, then evolved at the actual $Q^2$ using the DGLAP equations, and combined with the perturbative coefficient functions.

\section{DIS cross sections and PDFs}
The kinematics of the lepton-nucleon reaction is conveniently described by the variables $x$, $Q^2$ , already introduced, and $y$, the inelasticity, representing the energy transfer between the lepton and the hadron system. The inelasticity $y$ is directly related to the lepton-parton scattering angle, and can be used to study the helicity structure of the interaction.  The kinematic variables are related by $Q^2=s x y$, where $s$ is the lepton-proton centre-of-mass energy.
\par 
Neutral current (NC) and charged current (CC) processes are measured most recently by the H1 and ZEUS experiments in $e^\pm p$ collisions at HERA. The NC events contain a prominent electron  and a jet of particles measured in the calorimeter, while in CC events only the jet is visible since the outgoing neutrino in not detected. 
\par
Since a large domain in $x$ and $Q^2$ is accessed, the NC cross section become sensitive to weak effects, besides the photon exchange, which dominates the cross section. The $Z^0$ boson exchange can be incorporated into the so-called generalised structure functions. The cross section is parameterised as: 
\begin{equation}
\frac{\rm{d}^2\sigma^{\pm}_{NC}}{{\rm d}x{\rm d}Q^2}=
\frac{2\pi\alpha^2}{xQ^4}(Y_+\tilde{F}_2{\mp}Y_-x\tilde{F}_3-y^2\tilde{F}_L) 
\,\,\,,
\label{eq:ncxsec}
\end{equation}
The helicity dependence of the electroweak interactions is given by
the terms $Y_{\pm}=1\pm(1-y^2)$.
The generalised structure functions $\Ftwo$ and $x\Fz$ can be further
decomposed~\cite{klein} as:
\begin{eqnarray}
 \label{f2p}
\nonumber  \Ftwo  \equiv & \Fem & - \ v_e  \ \frac{\kappa  \QQ}{(\QQ + M_Z^2)}
  \Fint  \,\,\, + (v_e^2+a_e^2)  
 \left(\frac{\kappa  Q^2}{\QQ + M_Z^2}\right)^2 \Fwk\,, \\
\nonumber  \label{f3p}
 x\Fz    \equiv &      & - \ a_e  \ \frac{\kappa  \QQ}{(\QQ + M_Z^2)} 
 x\Fzint + \,\, (2 v_e a_e) \,\,
 \left(\frac{\kappa  Q^2}{\QQ + M_Z^2}\right)^2  x\Fzwk\,,
\end{eqnarray} 
with $\kappa^{-1}=4\frac{M_W^2}{M_Z^2}(1-\frac{M_W^2}{M_Z^2})$ in the
on-mass-shell scheme.  The quantities $v_e$ and $a_e$ are the
vector and axial-vector weak couplings of the electron or positron
 to the
$Z^{0}$.  The electromagnetic structure function $\Fem$
originates from photon exchange only
 and dominates in most of the accessible phase space.
\par
The functions $\Fwk$ and $x \Fzwk$
are the contributions to $\Ftwo$ and $x\Fz$ from $Z^0$ exchange and the
functions $\Fint$ and $x\Fzint$ are the contributions from $\gamma Z$
interference. These contributions are significant only at high $Q^2$. 
For longitudinally unpolarised lepton
beams the $\Ftwo$ contribution is the same for $e^-$ and for $e^+$ scattering,
while the $x \Fz$ contribution changes sign as can be seen in
eq.~\ref{eq:ncxsec}.
The longitudinal structure function $\FL$ may be
decomposed in a manner similar to $\Ftwo$. Its contribution is
significant only at high $y$.
\par
In the quark parton model (QPM) the structure functions $F_2$,
$F_2^{\gamma Z}$ and $F_2^Z$ are related to the sum of the quark
and anti-quark momentum distributions, $xq(x,Q^2)$ and $x\overline{q}(x,Q^2)$: 
\begin{equation}
 \label{eq:f2}
 [F_2,F_2^{\gamma Z},F_2^{Z}] = x \sum_q 
 [e_q^2, 2 e_q v_q, v_q^2+a_q^2] 
 \{q+\bq\} 
\end{equation}
and the structure functions $xF_3^{\gamma Z}$ and $xF_3^Z$ to the
difference, which determines the valence quark distributions, $xq_v(x,Q^2)$:
\begin{equation}
 \label{eq:xf3}
 [ x F_3^{\gamma Z},x F_3^{Z} ] = 2x \sum_q 
 [e_q a_q, v_q a_q]
 \{q -\bq \} = 2 x \sum_{q=u,d} [e_q a_q, v_q a_q] q_v\,.
\end{equation}
In equations~\ref{eq:f2} and \ref{eq:xf3}, 
$v_q$ and $a_q$ are the vector and
axial-vector weak coupling constants of the quarks to the $Z^0$, respectively. 
\begin{figure}
\centerline{
\epsfxsize=8.0cm\epsfysize=8.0cm\epsfbox{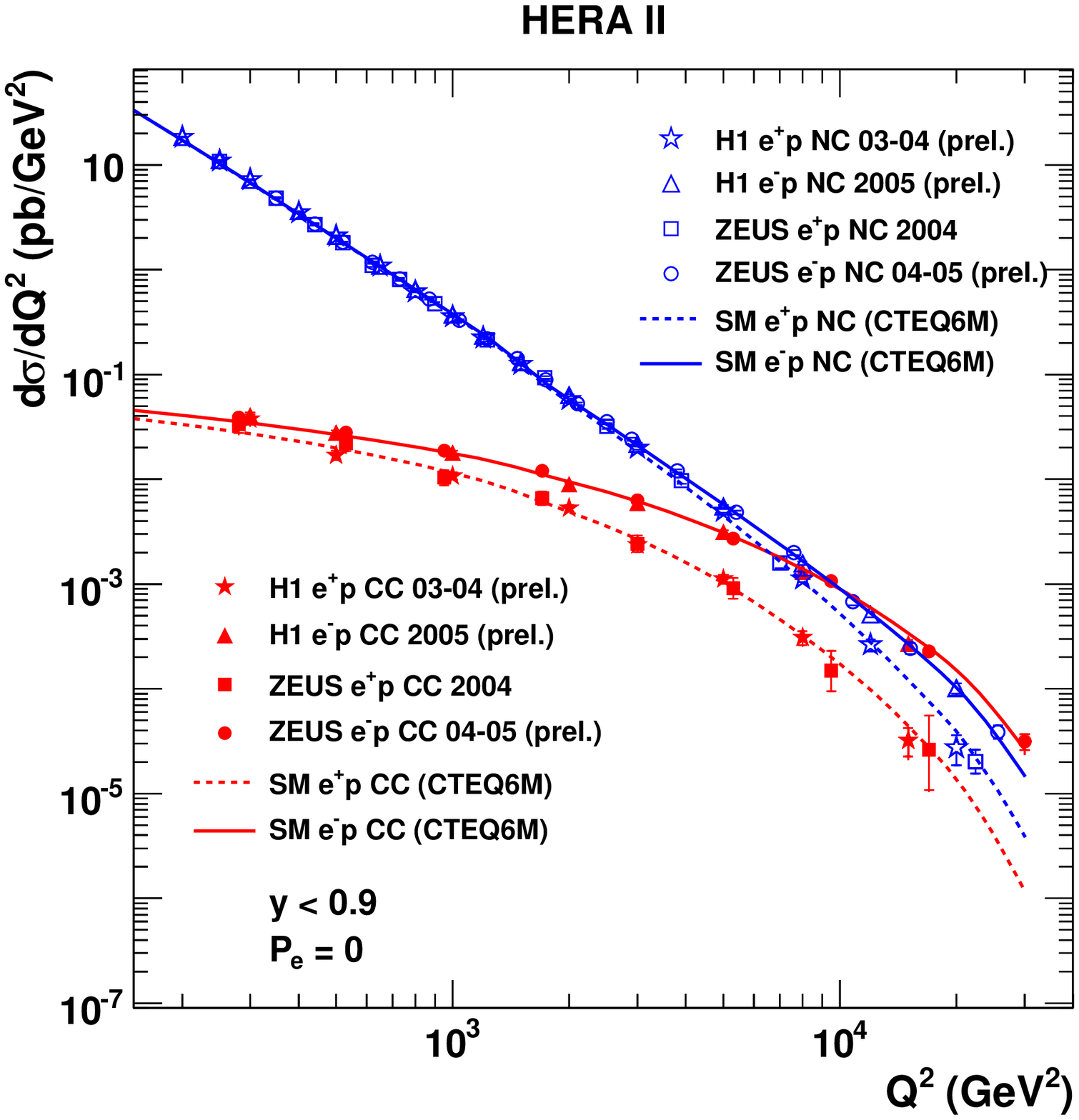}
\epsfxsize=8.5cm\epsfbox{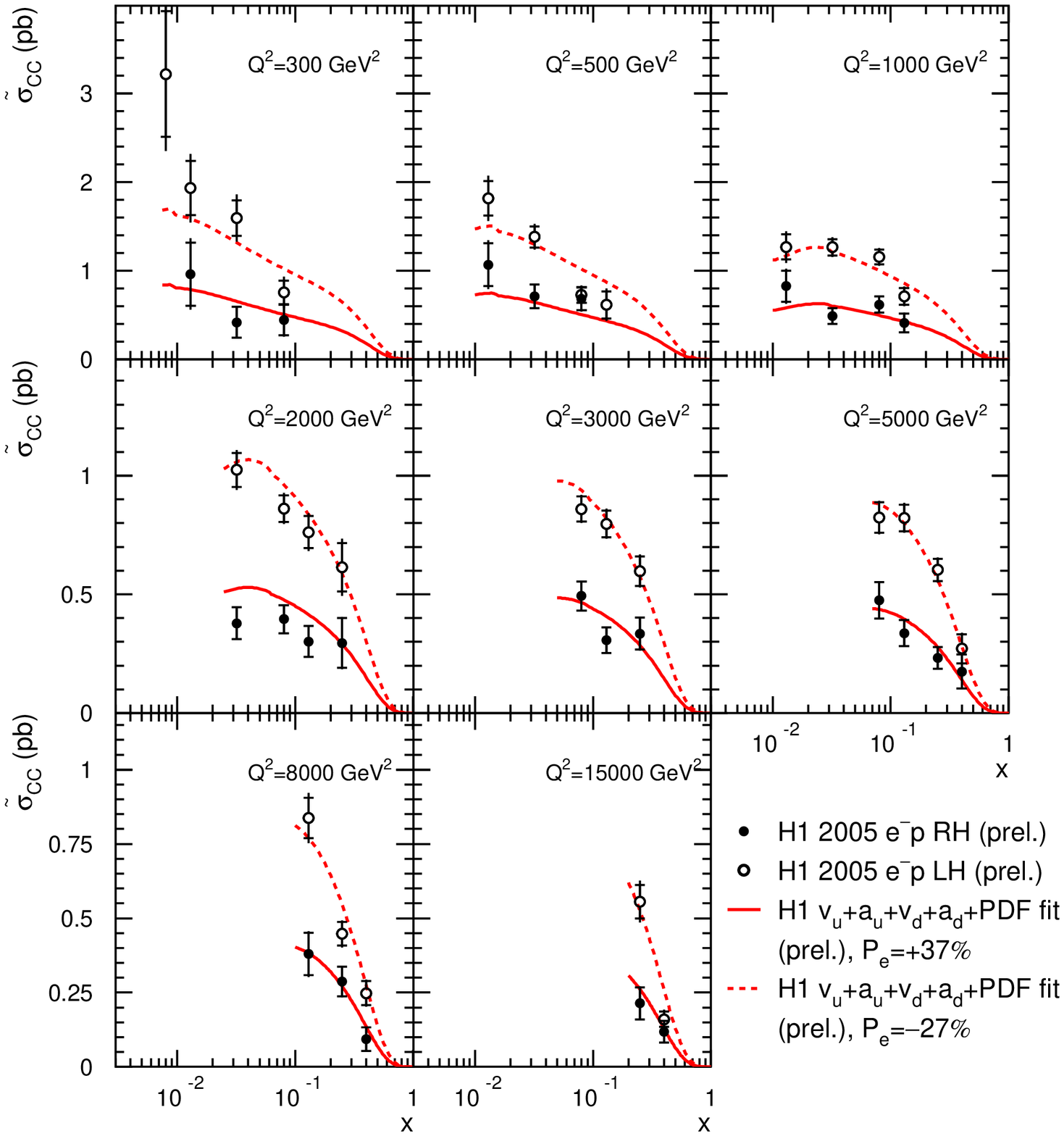} }
\caption{Left: the neutral current and charged current cross section as a function of $Q^2$ measured in $e^\pm p$ collisions at HERA. Right: the charged current reduced cross section $\tilde{\sigma}_{CC}$ as a function of $x$ for various $Q^2$ values measured in electron--proton collisions with different electron beam polarisations. }
\label{fig:ccnc}
\end{figure}
\par
The charged current (CC) interactions, 
$e^\pm p \rightarrow \overline{\nu}_e^{\mbox{\tiny
\hspace{-3mm}\raisebox{0.3mm}{(}\hspace{2.5mm}\raisebox{0.3mm}{)}}}X$,
are mediated by the exchange of a $W$ boson in the $t$ channel.  The cross section is parametrised as:
\begin{eqnarray}
\nonumber \frac{d^2\sigma^{\rm CC}(e^\pm p)}{dxdQ^2}&=&
\frac{G^2_F}{2\pi x}\left[\frac{M^2_W}{M^2_W+Q^2}\right]^2\tilde{\sigma}^\pm_{CC}(x,Q^2)
\,,\label{eqn:xscc}\\
\nonumber {\rm with} \hspace{5mm} \tilde{\sigma}^\pm_{CC}(x,Q^2)&=&
\frac{1}{2}\left[Y_+W_2^\pm (x,Q^2)\mp Y_-xW_3^\pm (x,Q^2)-y^2W_L^\pm (x,Q^2)\right]\,.
\end{eqnarray}
The reduced cross section is denoted by $\tilde{\sigma}$ is the reduced cross section, $G_F$ is the  Fermi constant, $M_W$, 
the mass of the $W$ boson, and $W_2$,
$xW_3$ and $W_L$, CC structure functions.
In the QPM 
the structure functions $W^\pm_2$ and $xW^\pm_3$ may be interpreted as
lepton charge dependent sums and differences of quark and anti-quark 
distributions:
\begin{equation}
\nonumber W^+_2=x(\overline{U}+D),\hspace{2mm} xW^+_3=x(D-\overline{U}),\hspace{2mm}
W^-_2=x(U+\overline{D}),\hspace{2mm} xW^-_3=x(U-\overline{D})\,,
\end{equation}
whereas $W^\pm_L=0$. 
The terms $xU$, $xD$, $x\overline{U}$ and $x\overline{D}$
are defined as the sum of up-type, of down-type and of their anti-quark-type 
distributions, i.e. below the $b$ quark mass threshold:
$xU=x(u+c) , \hspace{3mm}
xD=x(d+s),\hspace{3mm}
x\overline{U}= x(\overline{u}+\overline{c}),\hspace{3mm}$ and 
$x\overline{D}= x(\overline{d}+\overline{s})\,$.
\par
The differential NC and CC cross sections as a function of $Q^2$ are shown in figure~\ref{fig:ccnc}~(left) 
for $e^-p$ and $e^+p$ collisions at HERA. At low $Q^2$ the NC cross section, driven by the electromagnetic interaction, is two orders of magnitude larger than the CC cross section and corresponds to a pure weak interaction. At large $Q^2\sim M_{W,Z}^2$ the two cross sections are similar.
The largest $Q^2$ measurement corresponds to a resolution of $\delta~\simeq 10^{-18}$~m, i.e. 1/1000 of the proton size. The agreement between the measurement and the prediction based on QCD improved parton model suggests no evidence for quark substructure.
\par
The double differential reduced cross section $\tilde{\sigma}_{CC}(x,Q^2)$ is shown in figure~\ref{fig:ccnc}~(right). Two data sets are shown, with different polarisations of the electron beam. The enhancement of the cross section as a function of the beam polarisation $\sigma_P^{e\pm p} = (1\pm P_e) \sigma_{nopol}$ is visible.  
In addition, the CC processes are sensitive to individual quark flavours, which is especially visible at large $Q^2$: the $e^+p$ collisions probe the $d(x)$ quark distribution, while $e^-p$ are more sensitive to the $u(x)$. This is a very useful feature of the CC processes compared to the NC, where the flavour separation is weaker and occurs only at high $Q^2$ where the $Z$ boson exchange is important.

\section{H1 and ZEUS cross section combinations}

\begin{figure*}[t]
\centering
\includegraphics[width=80mm]{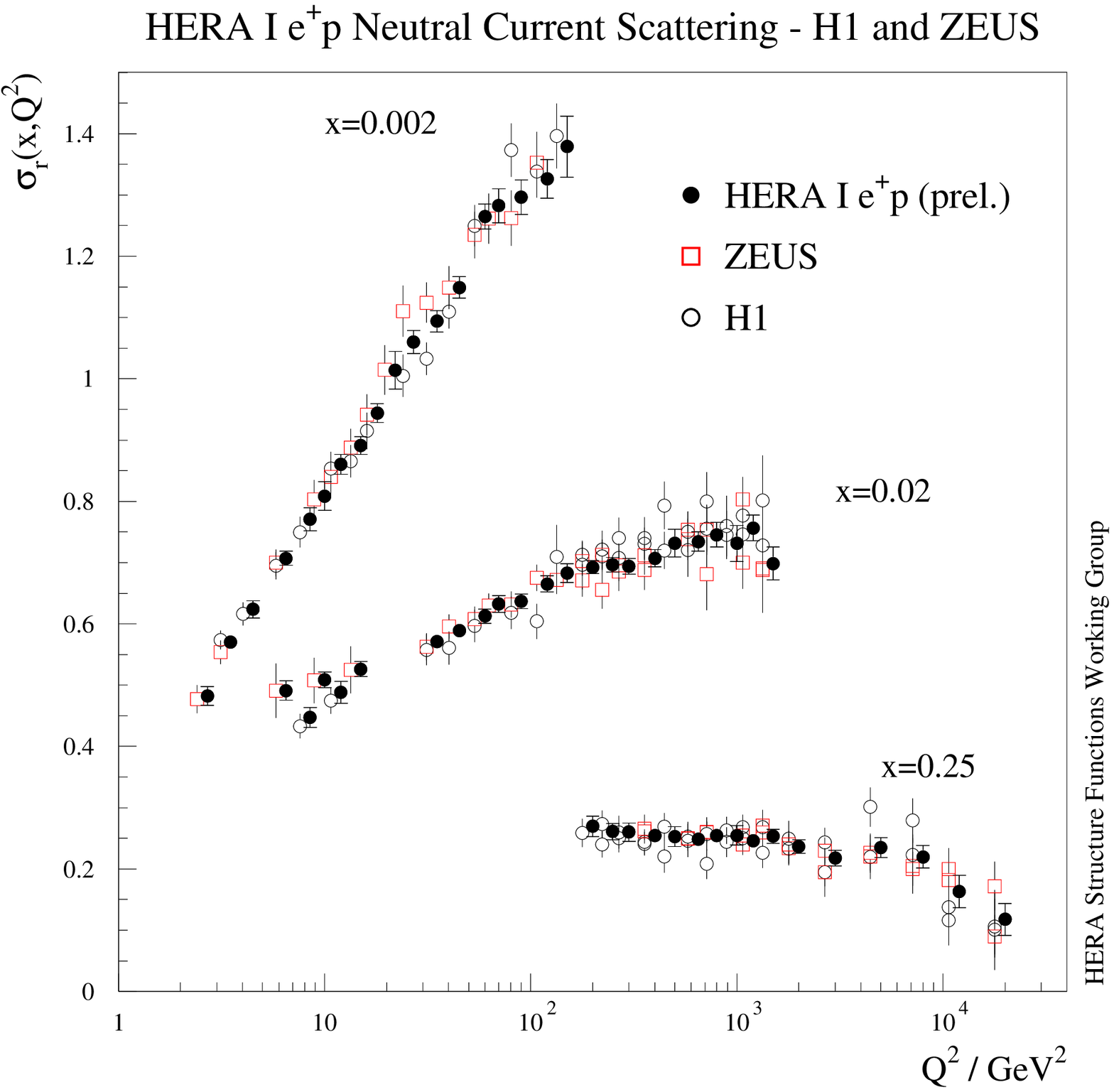}
\includegraphics[width=80mm]{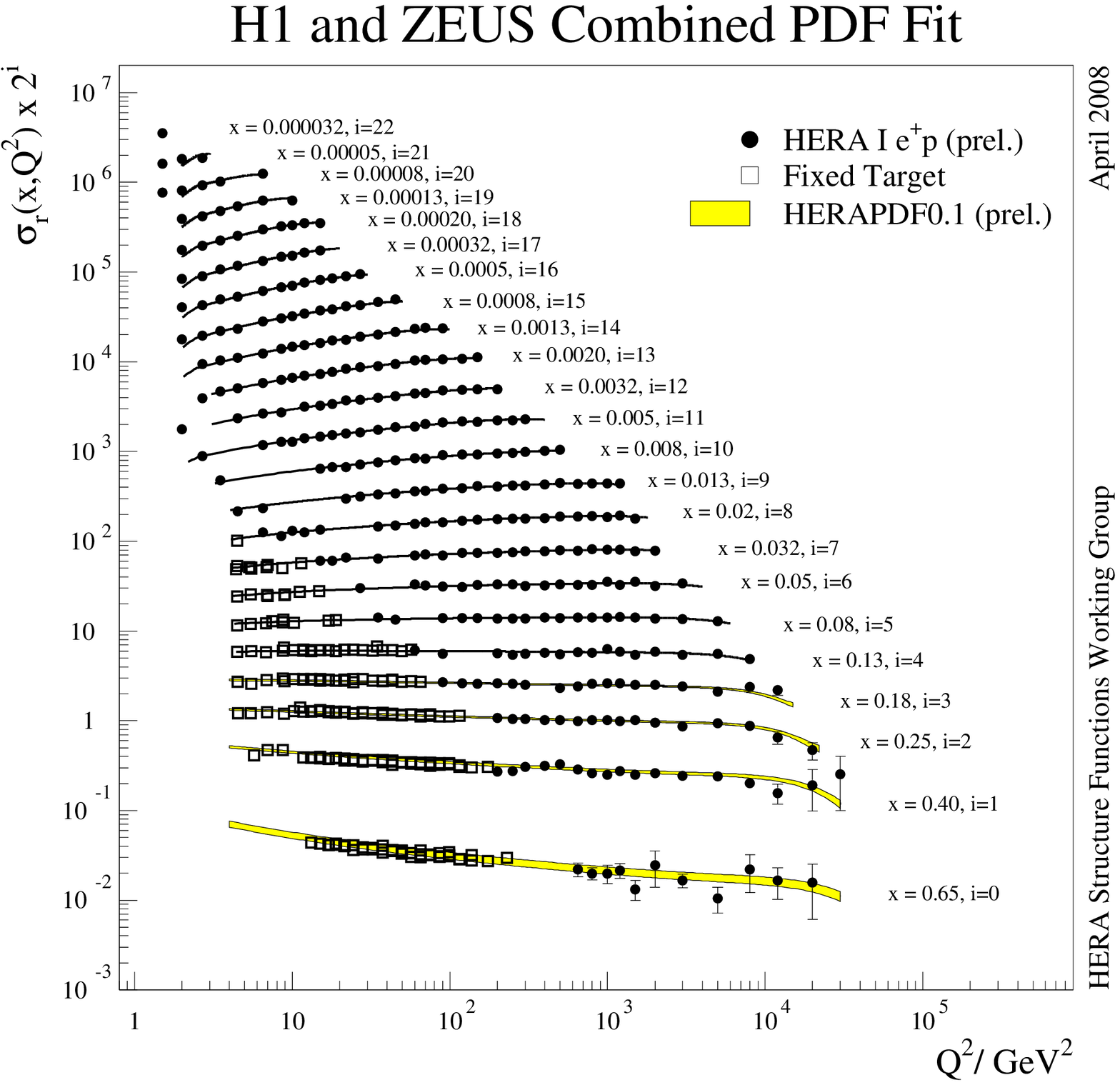}
\caption{Left: Examples of cross section combinations for three bins in $x$ as a function of $Q^2$. The precision of the combined experimental points is dramatically improved versus the individual measurements. Right: The combined HERA I data set compared with the HERAPDF 0.1 fit. } \label{f2_herapdf}
\end{figure*}

The cross section measurements are  generally affected by statistical and systematic uncertainties. The latter can also be correlated among various measurements, leading to common shifts or coherent change of shapes in the determined PDFs. These uncertainties and their correlations should be properly taken into account in the fit procedures, in order the propagate the experimental uncertainties of the PDFs determinations into predictions for cross sections, for instance at LHC. The fit procedure effectively combines the information of the same measurements (for instance the cross section in the same phase space or bin). However, this combination is done via the regular assumptions of such fits: input parameterisations, tolerances among different  data sets and so on.    Alternatively, the  measurements of the same observables can be combined in model independent way using  a coherent procedure~\cite{Glazov:2005rn} that takes into account the statistical and sytematic errors. The  main advantage of such procedure is the proper treatment of the correlated  and uncorrelated errors. The obtained data set therefore reflects the experimental knowledge in the given measurement phase space and is factorised from the fit procedures, which can be therefore better understood and optimised. 
\par
The average procedure is based on the minimisation of a $\chi^2$ function, whose free parameters are not only the average values in each measurement point but also the correlated systematics shifts from each data set. The minimisation procedure is designed to find the best estimation of the local average measurements, also taking into account also long range systematic shifts that may occur due to the correlated systematic errors of the measurements.
This averaging procedure achieve a significant reduction of the long-range correlated systematic errors, as illustrated in figure~\ref{f2_herapdf} for three sets of measurements at low values of $x$. The gain in precision is significant and better than the simple combination of errors in each point. This can be explained by the fact that for each measurement point a given correlated uncertainty of one set of measurements is constrained by the independent measurements from other data sets.  This cross-calibration leads to a precise and coherent HERA data set to be used to determine the PDFs.

\section{QCD fits}
 As explained above, a variety of data from DIS and hadronic collisions can be used to constrain the PDFs in the so-called global fits.  The approach is based on the factorisation assumption and provides PDF parameterisations that can be used for further predictions, for instance at LHC. The consistency of this procedure provides also a stringent test of the QCD. Recent progress has been reported in including Tevatron data and refining the error evaluations~\cite{Watt:2008hi} and in detailed studies of the precision of the observables and its correlation with the PDFs~\cite{Nadolsky:2008zw}. Such approaches are the main source for PDFs parameterisations since they incorporate all available information on the proton structure.
\begin{figure*}[t]
\centering
\includegraphics[width=78mm]{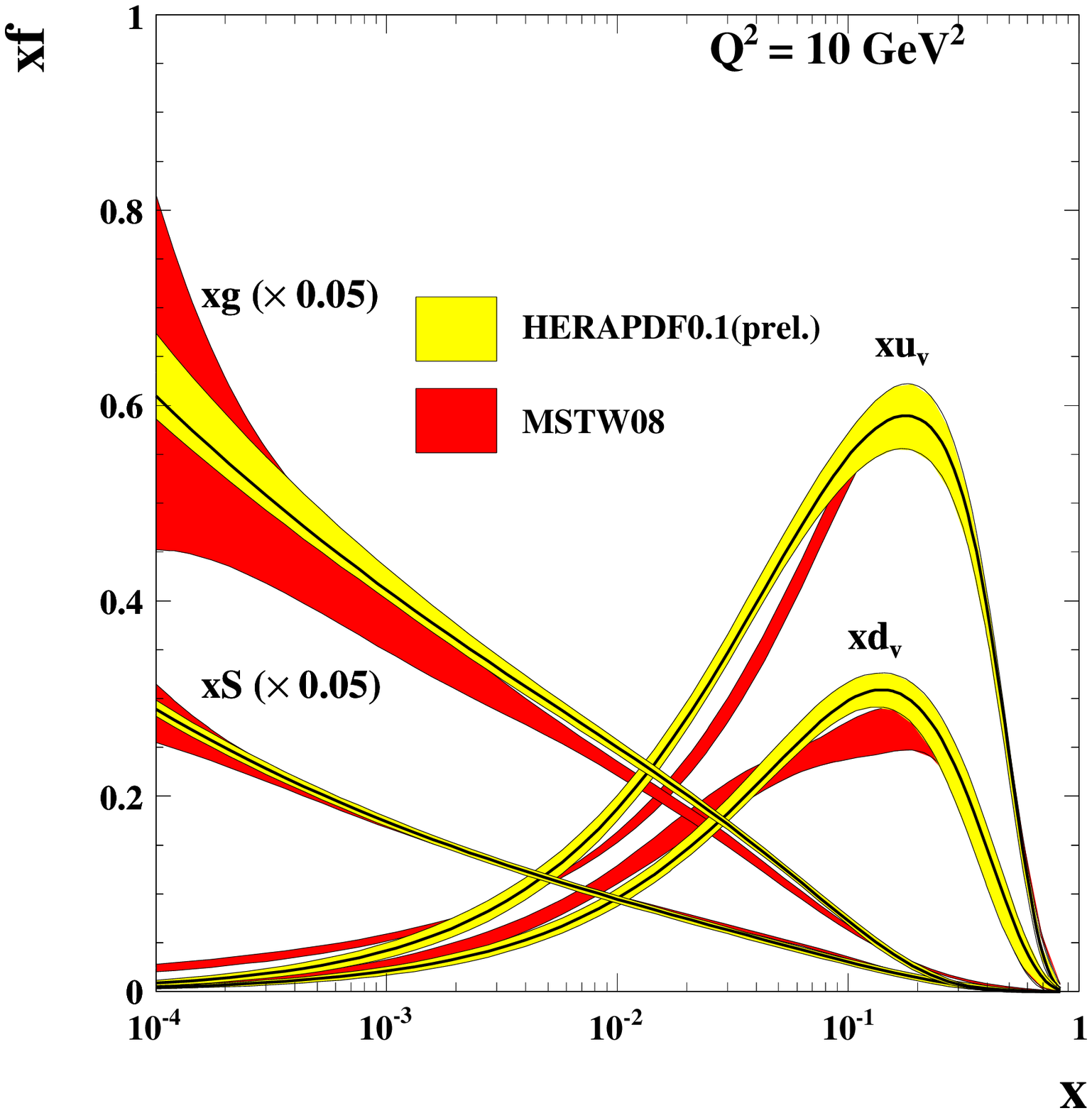} \hspace{1cm}
\includegraphics[width=78mm]{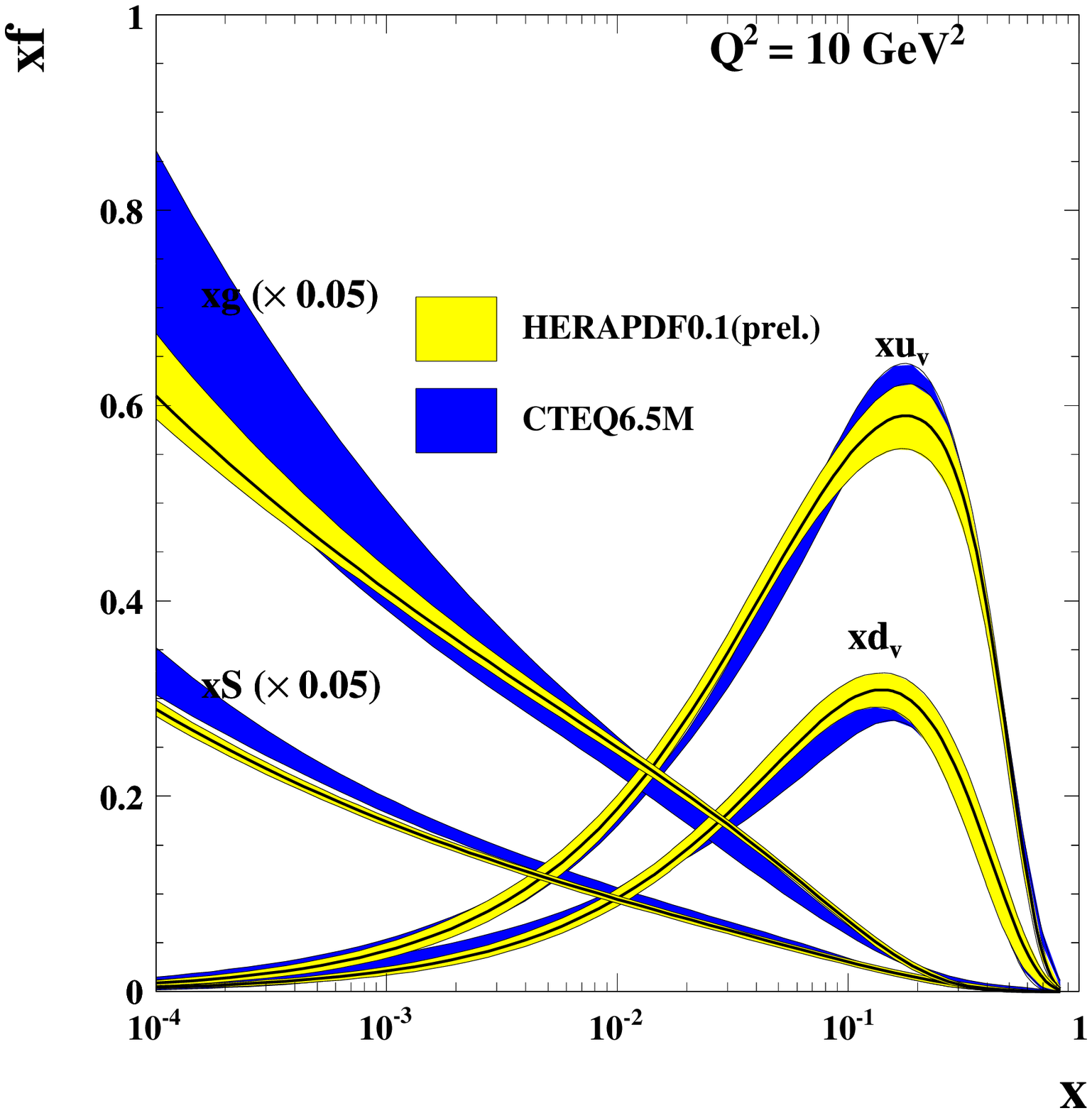}
\caption{HERAPDF0.1 fit compared with MSTW and CTEQ fits.}
\label{herapdf_gfits}
\end{figure*}
\begin{figure*}[t]
\centering
\includegraphics[width=170mm]{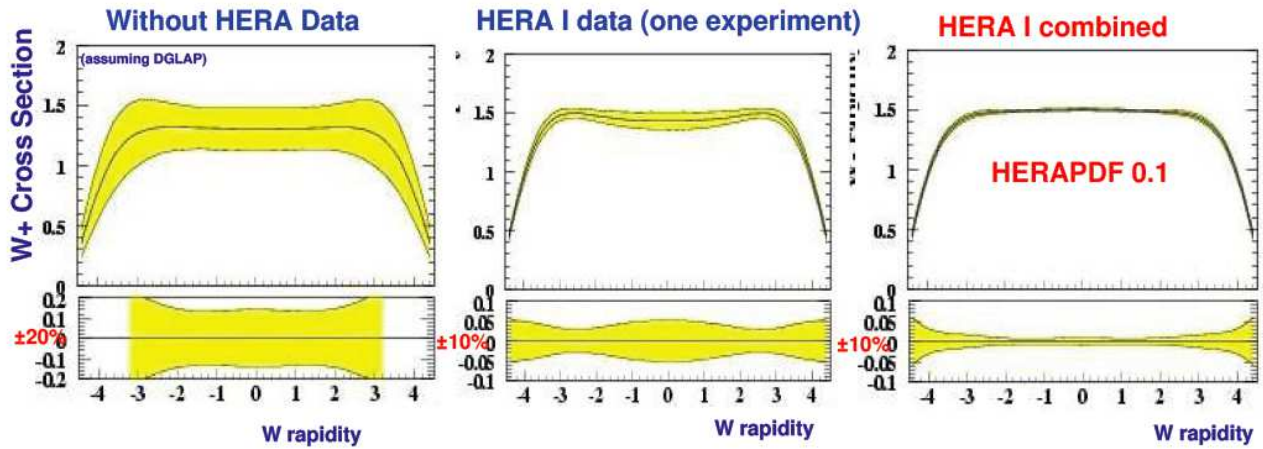}
\caption{The improvement in precision obtained using HERA data: prior to HERA data (left), using one HERA experiment (center) and using HERAPDF~0.1 parametrisation (right). Note that
pre-HERA data do not constrain W production at LHC and the precision shown
in the left figure fully relies on parameterisation extrapolation from large $x$ measurements from fixed target experiments. }
\label{herapdf_wlhc}
\end{figure*}
 \par
However, it is well known that the global fits encouter and reveal tensions among various experimental data sets and, in absence of further experimental information, approximate statistical methods are used in order to cope with these inconsistencies. The most striking issue in the global fit procedures is the estimation of PDF errors, which has to be done with an somewhat arbitrary prescription of $\Delta \chi^2\simeq 50\div 100$ in order to accomodate systematic differences between various data sets. The PDF parametrisations obtained in the global fits using various approaches are  in good agreement most of the time, but there are also some cases where the predictions differ by more than the quoted PDF uncertainty~\cite{Djouadi:2003jg, Adam:2008ge}.  In addition, the various improvements of the treatment of the theoretical input have lead in the past to variations of the predicted cross sections at LHC by significantly more than the previously cited uncertainties~\cite{Tung:2006tb}. This situation raises the question of the theoretical errors associated with the extracted PDFs.
 \par
 In these conditions, a useful approach is to use one coherent data set covering a significant area of the phase space, in order to study the QCD fit behavior, to compare to other data sets and to calculate predictions for observables at LHC.
The newly obtained HERA combined data set has the advantage of very small systematic errors, which in fact no longer dominate. This makes the QCD fit procedure more robust and allows a test of the influence of the theoretical assumptions.
The HERA combined preliminary data, described above, has been used by the H1 and ZEUS structure functions working group to perform a stand alone NLO QCD fit,  HERAPDF~0.1~\cite{herapdf}.   
The consistency of the input data set simplifies the determination of the experimental uncertainties on the PDFs, performed using the more standard tolerance condition $\Delta \chi^2=1$. 
\par
In figure~\ref{f2_herapdf} (right), the NC DIS combined reduced cross section is compared to the HERAPDF~0.1 preliminary fit. The impressive precision of the combined data is reflected in the precision of the fit, which, extrapolated to lower $Q^2$, is found to be in very good agreement with the fixed target DIS data. The precision of the new data lead to precise PDFs, in particular at low $x$, where most of the cross-calibration that occurs during the averaging procedure  dramatically improves the systematic errors of the combined cross section. The obtained parameterisations together with the estimated errors are compared in  figure~\ref{herapdf_gfits} to the global fits based on separate data sets. 
\par
Although the observed improvement in precision is mostly due to the improved combined data, the fit procedure itself, which is similar to the global fits, contains an intrinsic artificial constraint of the estimated error due to the functional form assumed for the PDFs at the initial scale $Q^2_0$. Recent progress in the field of PDF parametrisation opens the way to an model independent assesement of the PDFs~\cite{DelDebbio:2004qj}.
\par
The parametrisation HERAPDF~0.1 is used to predict the $W$ boson production in $pp$ collisions at LHC. The improvement of the precision of this observable is shown in figure~\ref{herapdf_wlhc}. The figure~\cite{mandy} displays the unique role of HERA for LHC physics and the significant gain in precision obtained recently from the H1 and ZEUS combined data. More HERA data, in particular at high $x$ and $Q^2$  and from high statistics jet production analysis, are expected in the near future. They will further improve the precision of PDFs and are therefore an essential input for the LHC physics.

\section{The direct measurement of the structure function $F_L$ at low $x$}

The longitudinal structure function  $F_L$ is a fundamental form factor of the proton, as can be seen from equation~\ref{eq:ncxsec}. 
Its contribution to the cross section is usually  small, and only visible at large $y$ where the weighting factor approaches the unity. 
Nevertheless, the structure function $F_L$ is of particular importance.
In the naive quark-parton model, the longitudinal structure function $F_L =F_2-2xF_1 \equiv 0$ (the Callan-Gross relation) reflecting the fact that longitudinal photons with helicity $\pm1$ cannot absorb spin $1/2$ quarks. 
This is possible only if a gluon radiation allows a spin flip of the interacting quarks.
It can be shown that $F_L$ is directly related to the gluon density in the proton~\cite{Altarelli:1978tq,Cooper-Sarkar:1987ds} $xg(x) = 1.8[ \frac{3 \pi}{2 \alpha_s} F_L(0.4 x) - F_2(0.8 x)] \simeq \frac{8.3}{\alpha_s} F_L$  meaning that at low $x$, to a good approximation $F_L$ is a direct measure for the  gluon distribution.
\par
\begin{figure*}[t]
\centering
\includegraphics[width=135mm]{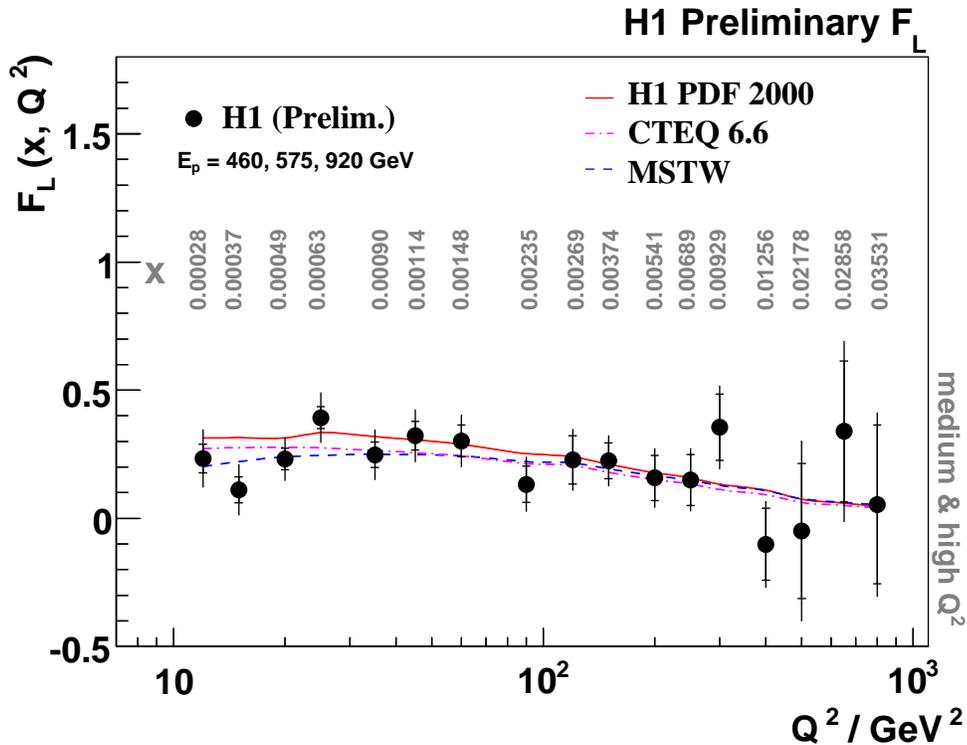}
\caption{The measurement of the longitudinal structure function $F_L$ at HERA.}
\label{fl}
\end{figure*}

A direct measurement of $F_L$ can be performed if the cross section $\sigma (E_p) \sim F_2 (x,Q^2) + f(y) \; F_L  (x,Q^2) $ is measured at fixed $x$ and $Q^2$ but variable $y$. This can only be performed if the collision energy $\sqrt{s}$ is varied, for instance by reducing the proton beam energy at HERA. Then $F_L(x,Q^2)$ can be directly measured with reduced uncertainties from the difference of cross sections: $\displaystyle F_L \sim C(y)*(\sigma(E_p^1) -\sigma(E_p^2)) $.
 The data taking at lower proton energies took place at the end of the HERA  run in 2007. The data collected at proton energies of 460, 575 and 920~GeV have been used to perform the first direct measurement of $F_L$ in the low $x$ regime. The first measurements in the medium $Q^2$ range from 12~GeV$^2$ to 90~GeV$^2$  were published recently by the H1 collaboration~\cite{h1flmed}. 
Preliminary measurements have been performed by  H1~\cite{h1flq} in an extended range of $Q^2$ up to 800~GeV$^2$. The H1 measurements averaged in $x$ for each bin in $Q^2$ are shown in figure~\ref{fl}. The ZEUS collaboration  presented measurements in the $Q^2$ range from $24$ to $110$~GeV$^2$~\cite{zfl}. 
The measured $F_L$ is significantly different from $0$ at lowest $Q^2$ and are in agreement with the theoretical predictions based on various PDF parameterisations. It should be noted that in the presented phase space the gluon density and therefore the $F_L$ predictions are well (but indirectly) constrained by other (inclusive) measurements. The measurements at lower $Q^2<12$~GeV$^2$ are experimentally challenging, but will be explored soon at HERA. They will shed light on various theoretical approaches predicting different values for $F_L$ at low $x$~\cite{thorne} and together with other measurement in this kinematic regime will constraint even further the gluon component of the proton.

\section{Constraints from measurements in $p\bar{p}$ collisions  at Tevatron}
Precise measurements of cross sections in $p\bar{p}$ collisions for processes well understood theoretically, can also be used to constrain the PDFs, in particular in the high $x$ regime. Two such recent measurements are described here: the weak boson and jet production.   
\par
\begin{figure*}[t]
\centering
\includegraphics[width=95mm]{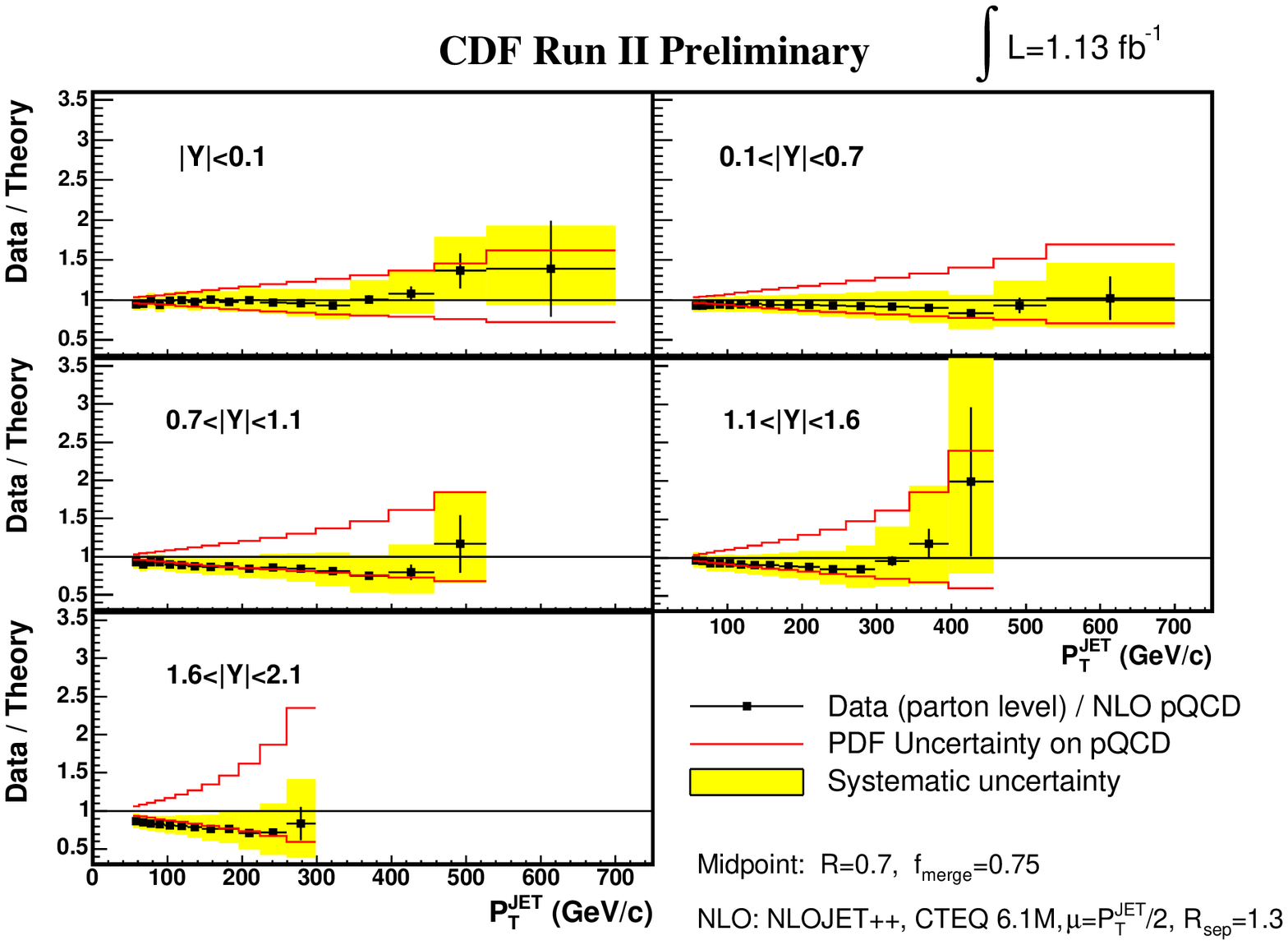}
\includegraphics[width=75mm]{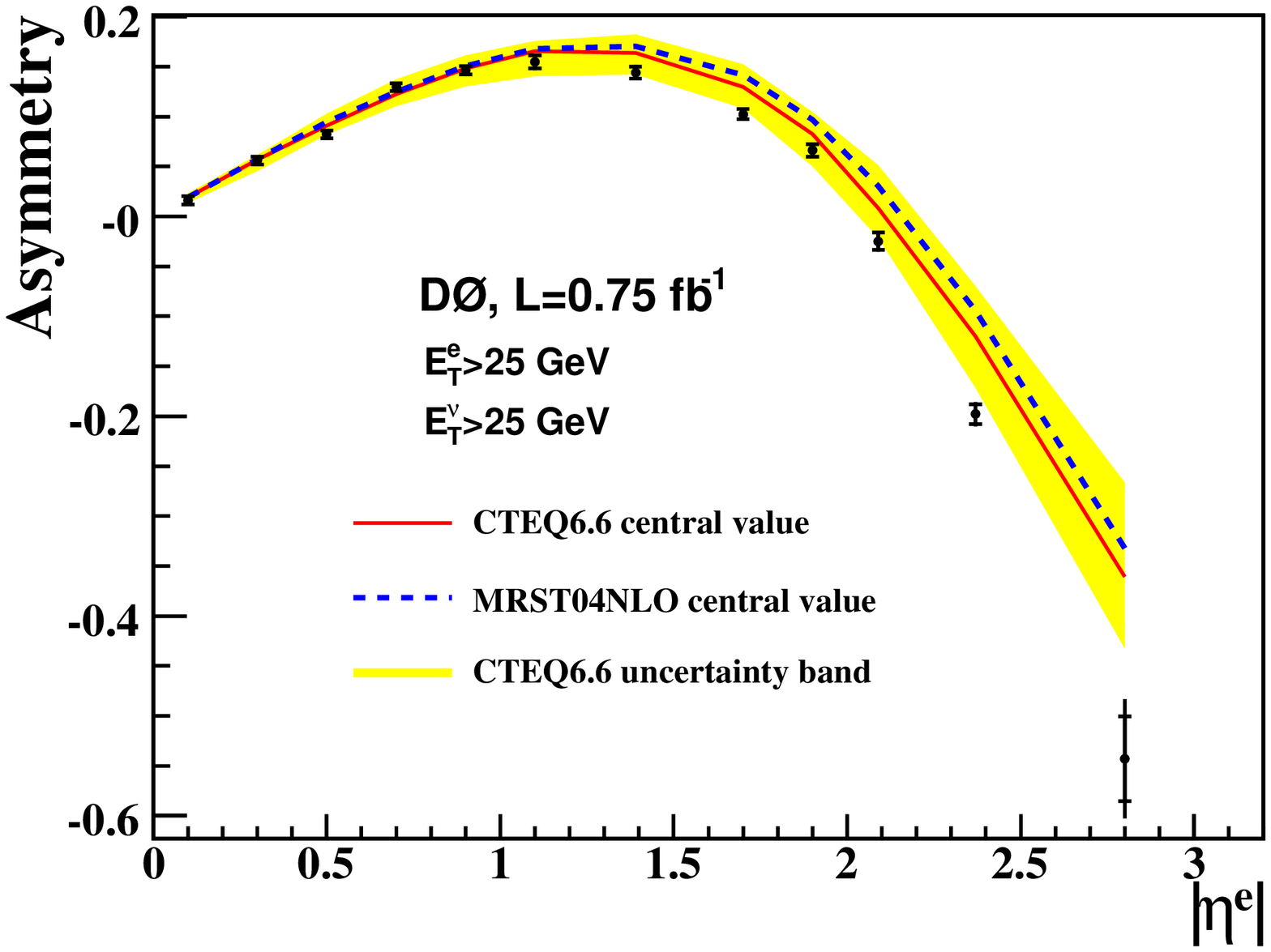}
\caption{Left: Jet production cross section. Right: W production asymmetry.} \label{tev}
\end{figure*}

Events with jets at Tevatron are produced by colliding quarks and gluons with high values of $x$. A recent measurement of the inclusive jet production was performed using a large statistics sample of CDF run II data~\cite{Aaltonen:2008eq}, corresponding to 1.1~fb$^{-1}$, using the cone-based midpoint jet clustering algorithm in the rapidity region of $|y|<2.1$ for jets with transverse momenta in the range from 60~GeV to 700~GeV. A similar measurement, using a $k_T$ jet clustering algorithm, was published by the D0 collaboration~\cite{d0:2008hua} for jet rapidities in the range -2.4 to 2.4 in the transverse momentum range from 50~GeV to 600~GeV. The results are consistent with next-to-leading-order perturbative QCD predictions based on recent parton distribution functions. The dominant experimental uncertainty is related to the hadronic energy scale. Cross sections are measured as a function of the jet transverse momentum and in various regions of rapidity. The measurements are in good agreement with the NLO QCD prediction, as can be seen in figure~\ref{tev}, which shows results from D0. The main theoretical uncertainties originate from the PDFs and exceed the experimental precision.  

\par

The measurement of  $W$ boson production at Tevatron also provides a probe of the proton structure. $W^\pm$ bosons are produced in $p\bar{p}$ collisions predominantly by the fusion of $u(d)$ quarks from the proton and $\bar{d}$($\bar{u}$) quarks from the anti-proton. As a simple example of the kinematics probed in this reaction at Tevatron and to compare with LHC, it can be noted that in order to produce a $W$ boson at rest by the fusion of a $u$ quark from the proton with an $\bar{d}$ quark from the antiproton, the two quarks should have $x=0.2$, while in the same configuration, the $W$ will be produced at rest at LHC by partons with $x=0.005$.  
\par
The $W^+$ or $W^-$ rapidity distributions reflect the quark momentum distribution inside the proton.   As the u quark tends to carry a higher fraction of the proton's momentum than the d quark, the $W^+$($W^-$) is boosted, on average, in the proton (anti-proton) direction. An asymmetry can be defined such that the charge of the $W$ boson is correlated to the proton or anti-proton beam directions. The $W$ charge asymmetry is defined as 
\begin{equation}
A(y_W) = (d\sigma(W+^)/dy_W - d\sigma(W^-)/dy_W)/ (d\sigma(W^+)/dy_W + d\sigma(W^-)/dy_W), 
\end{equation}
where $y_W$ is the rapidity of the W bosons. In the leading-order parton model, $A(y_W)$ is given approximately by
\begin{equation} 
A(y_W)= [ u(x1)d(x2) - d(x1)u(x2)]/[u(x1)d(x2) + d(x1)u(x2)] 
\end{equation}
 where $x_{1,2} = x_0e^{\pm yW}$ and $x_0 = M_W/\sqrt{s}$. Since the $W$ four-vector  cannot be unambigously reconstructed, a reweighting procedure is performed for the two reconstruction solutions obtained in each event by imposing the $W$ mass constraint. The measurements were performed recently by both D0 and CDF experiments~\cite{d0w,cdfw}, each using each more than 1~fb$^{-1}$ of run II data . The result obtained by D0 is shown in figure~\ref{tev}(right). The lepton charge asymmetry in $Z^0$ boson production leads also, in a similar way, to constraints on PDFs and has been measured recently with good precision at Tevatron~\cite{cdfdrellyan}. 
\par
The data from Tevatron measurements of the inclusive jets and $W$ and $Z$ charge asymetries were used in the most recent MSTW global fit  and lead to significant improvements, in particular for the gluon in the large $x$ domain~\cite{Watt:2008hi}. 


\section{Charm and beauty from  the proton}

\begin{figure*}[t]
\centering
\includegraphics[width=73mm]{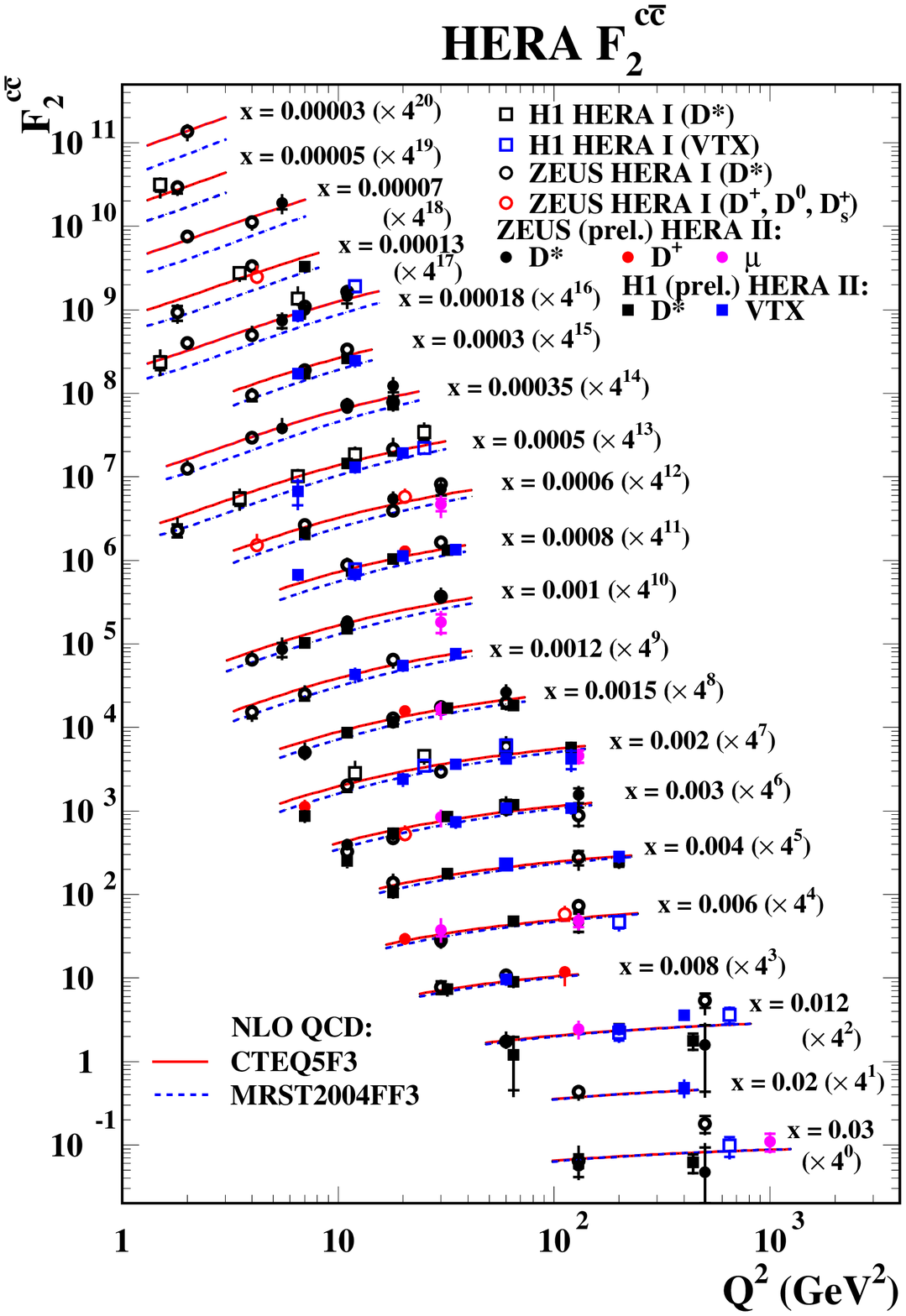}
\includegraphics[width=93mm]{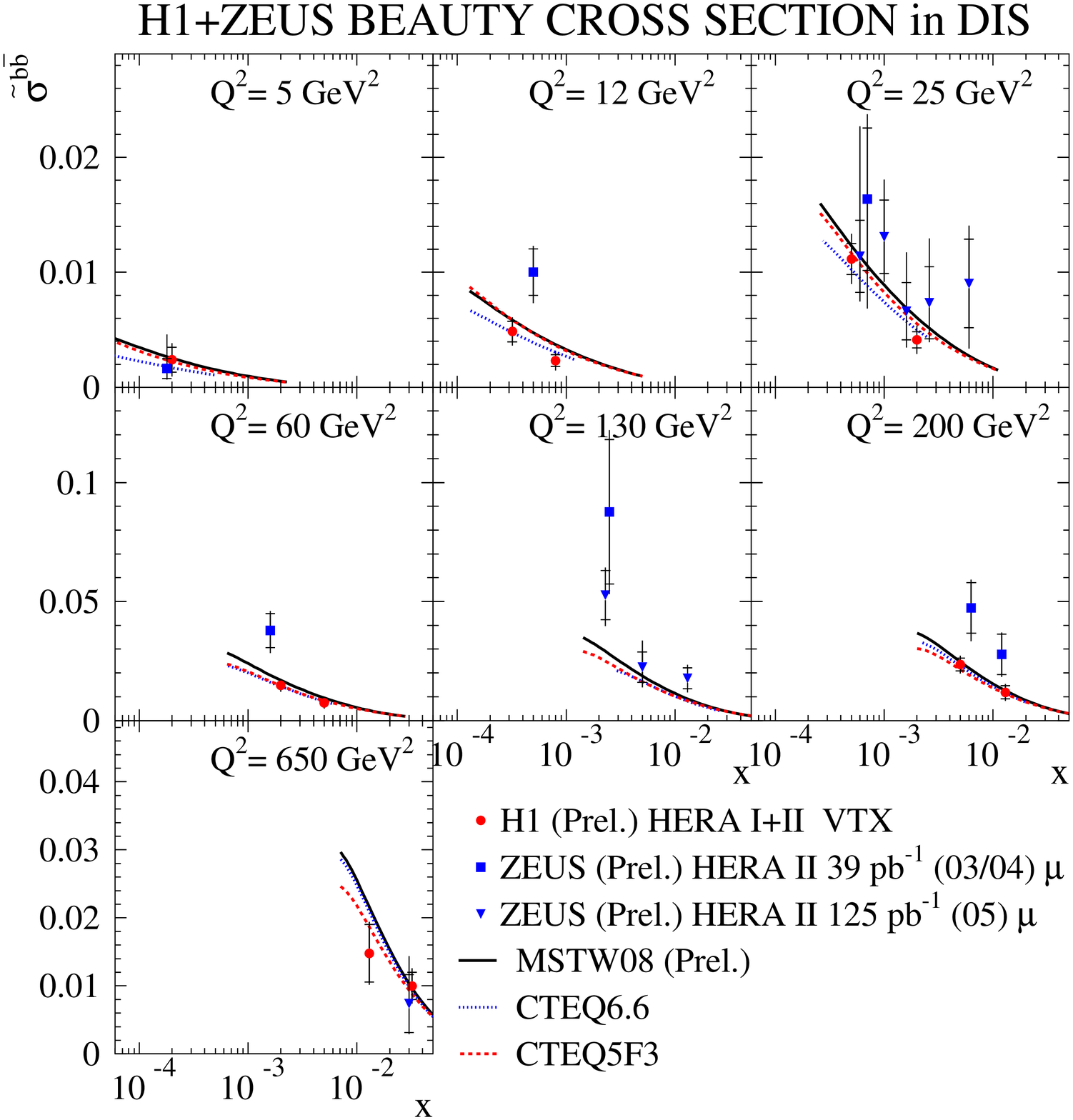}
\caption{The charm structure function (left) and the reduced cross section for beauty production (right) measurements at HERA compared to theorectical prediction. The precision of the data, in particular that of the new measurements, is comparable or better with the difference among the various predictions and may lead to improvements in the proton structure function determination.}
\label{sigma_b}
\end{figure*}
Of  particular importance for the search of particles that couple to the mass at the LHC, especially for the Higgs studies, is the understanding of the heavy flavour content of the proton. In DIS, the interactions initiated by heavy quarks can be tagged by the presence of  heavy flavour signals in the final state. Such interactions occur when gluons in the proton fluctuate to a pair of heavy quarks, one of which interacts with the photon (boson-gluon fusion), and therefore the study of  heavy flavour interactions in DIS may also constrain the gluon content of the proton. The cross section for charmed tagged events can be written in a similar way as the inclusive DIS cross section, as expressed in equation \ref{eq:ncxsec}. The main contribution to the charm (beauty) cross section originates from the structure function  $F_2^c$ ($F_2^b$). At leading order, the  $F_2^c$  and $F_2^b$ structure functions are proportional to the momentum distribution of $c$ and $b$ quarks inside the proton. 
  \par
  Measurements of heavy flavour production in DIS are performed at HERA~\cite{Kruger:2008cd} using two established experimental techniques: the tagging of $D^*$ mesons and the detection of particles emitted from long lived heavy hadrons. The measurements have been recently performed using the full available statistics from the HERA II run by the H1 collaboration. The production of $D^*$ mesons has been mesured in the kinematic range of $Q^2$ from 5~GeV$^2$ to 1000~GeV$^2$~\cite{h1dstar}. The measurements are in good agreement with the calculations based on NLO QCD and with LO Monte Carlo programs. An extrapolation of the measured cross section from the visible phase space to the full phase space is needed in order to extract the structure functions and to constrain the heavy quarks PDFs. The extrapolation is typically larger than a factor two and is dependent on theoretical assumptions. Therefore, a consistent theoretical framework has to be used in order to obtain the structure functions. The $F_2^c$ structure function based on those precise measurements has been extracted recently~\cite{h1f2cc}. 
\par
The measurement of processes with heavy quarks in the final state based on their lifetime use precise silicon tracking devices in order to identify tracks with significant distance of closest approach to the beam axis, potentially originating from heavy quarks decays. A recent measurement by the H1 collaboration~\cite{h1lifetime} uses a multi-variate analysis and exploits the full data sets collected at HERA. A good agreement is obtained with the theoretical calculations, while the extrapolation factors to obtain the structure functions are closer to unity than in the case of $D^*$ method. 
\par
A summary of the present status of the extraction of the heavy flavours structure functions $F_2^c$ and $F_2^b$ is presented in figure~\ref{sigma_b}. The obtained precision is comparable to the theoretical calculation.
The full potential of the HERA data will be obtained from the combination of H1 and ZEUS final data and may play an important role for contraints on the gluon and on the heavy flavour content of the proton. 
  \par 
  Other measurements directly test the flavour aspects of the proton interactions. The measurement of events containing strange particles in the final state were performed at HERA~\cite{Collaboration:2008ck,Chekanov:2006wz}. These measurements provide a new test of the QCD and allow a thorough investigations of strangeness production models. The link to the strange content of the proton is however difficult to establish since in the majority of events the strange particles are produced via the fragmentation process. Recent measurement at Tevatron~\cite{wchjet:2007dm} may also shed light on the strange content of the proton. The associated production of a $W$ boson and a charm particle has been measured using soft muon tagging in a data sample corresponding to an integrated luminosity of 1.8~fb$^{-1}$. The measured process is described at leading order by the scattering of a gluon predominanlty with a strange quark. The measurement is found to be in good agreement with the theoretical prediction, but is so far statistically limited. A precision of about 15\%  on the strange quark content of the proton can be achieved with  $\sim 6 $~fb$^{-1}$ of data, expected by the end of Tevatron run II.
  
\section{Study of the nucleon spin in polarised \boldmath{$ep$} collisions}
The understanding of the nucleon structure cannot be complete without the understanding of the mechanism by which the quarks and the gluons compose the spin of the nucleon $S_z=1/2$. The various contributions to the nucleon spin can be expressed as following: 
\begin{equation}\label{spinsum}
S_z=\frac{1}{2} = \frac{1}{2} \Delta \Sigma(\mu^2) + \Delta g(\mu^2) 
+ L_z^q(\mu^2) + L_z^g(\mu^2)\,.
\end{equation}
Here $\Delta \Sigma$  ($\Delta g$) describes the  integrated 
contribution of quark and anti-quark (gluon) helicities to the nucleon 
helicity and $L_z^q$  ($L_z^g$) is the $z$ component of  the orbital 
angular momentum among all quarks (gluons) at a given scale $\mu^2$. 
The main puzzle has been the observation that, contrary to the naive expectation, the quark contribution does not account for the nucleon spin. 
\begin{figure}
\centerline{
\epsfxsize=5.8cm\epsfbox{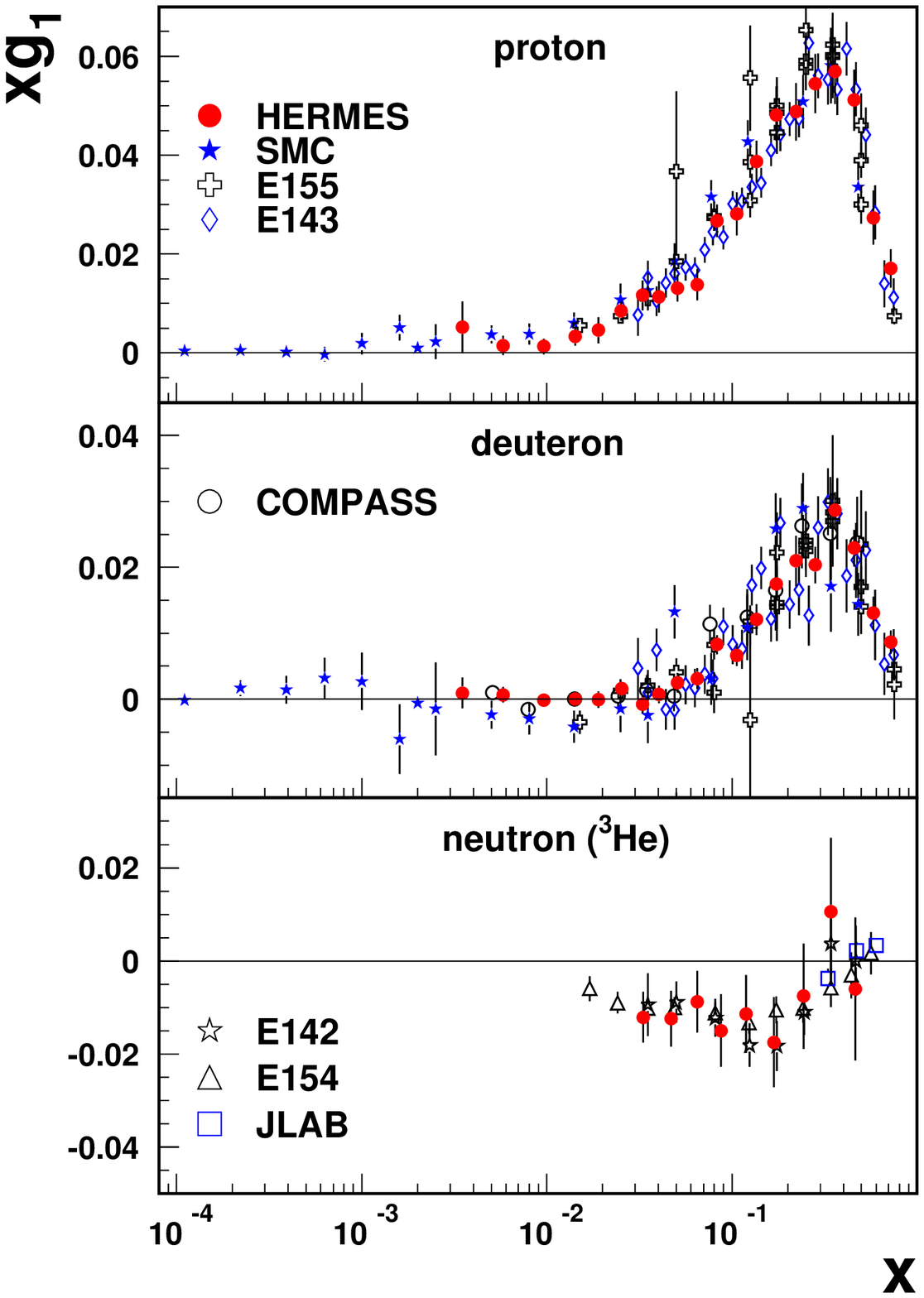}
\epsfxsize=5.8cm\epsfbox{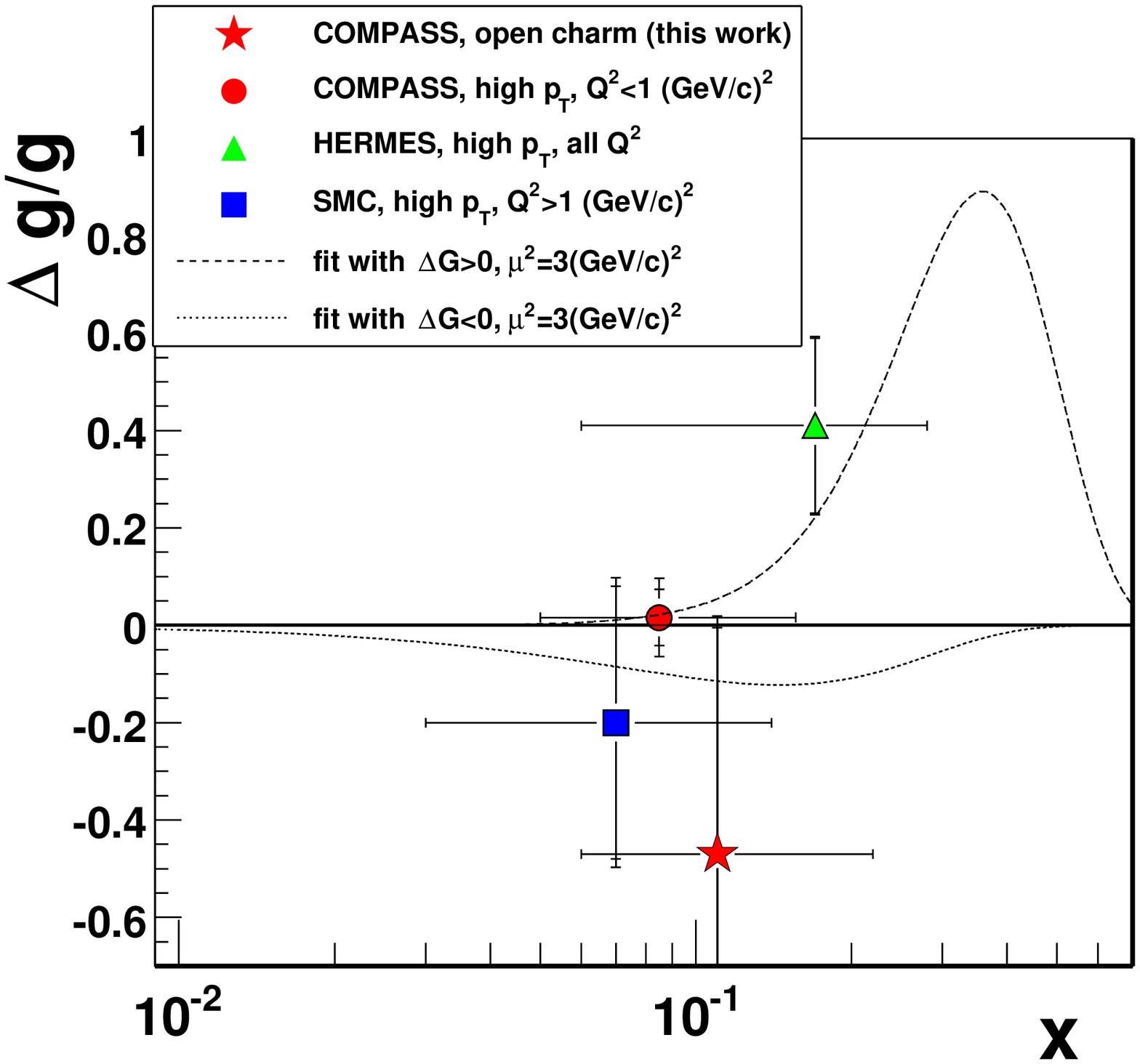}
\epsfxsize=7.4cm\epsfbox{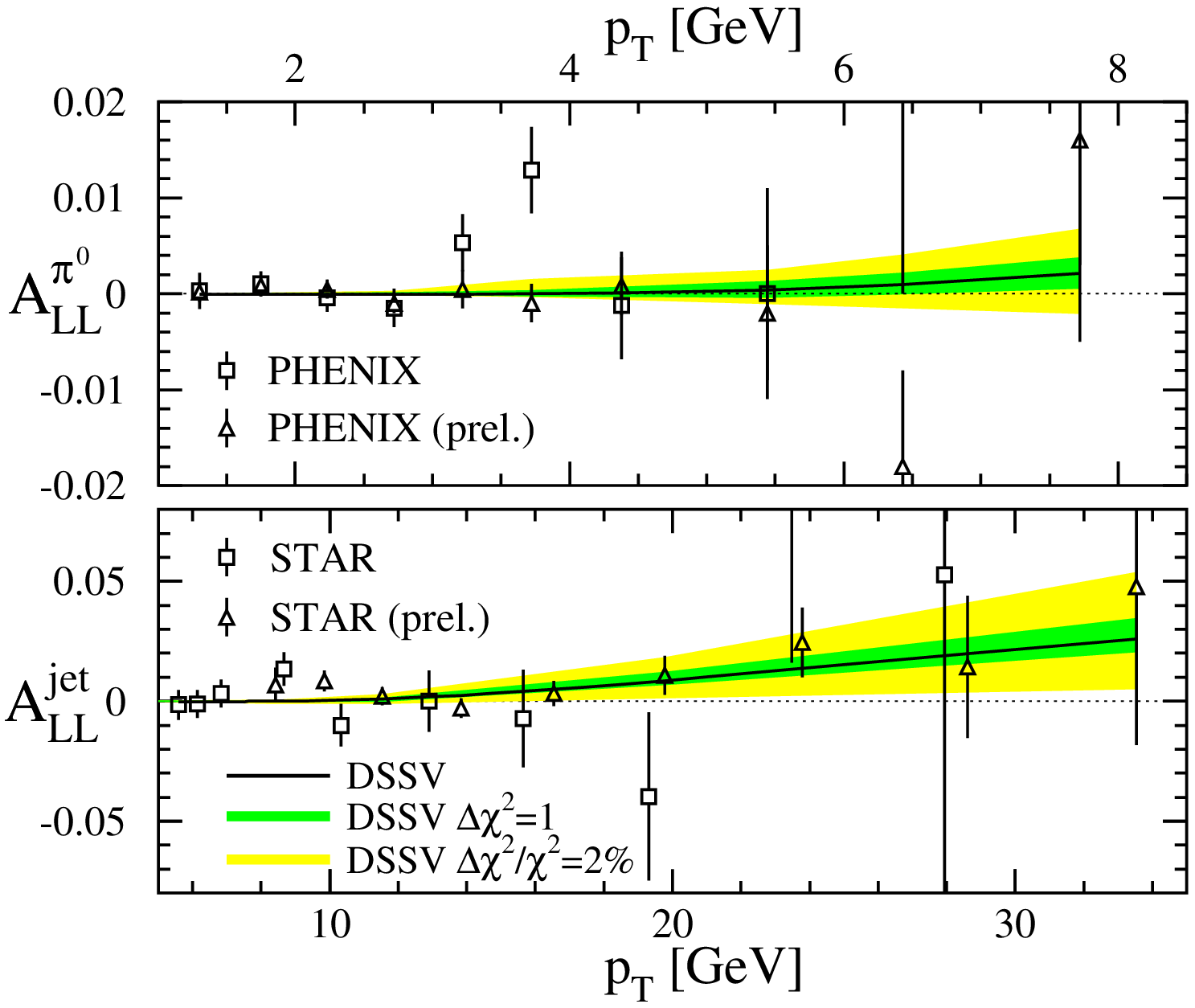}
}
\caption{Left: The measured structure function $g_1$ for proton, deuteron and neutron. The precise measurement from HERMES is compared with  measurements from other experiments. Center: The estimations of the gluon contributions $\Delta g/g$  (from~\cite{Alekseev:2008cz}). Right: Cross sections spin asymmetries for $\pi^0$ and jet production in polarised $pp$ collisions at RHIC, compared with a global fit.}
\label{fig:g1}
\end{figure}
\par
The spin-dependent DIS cross section can be parametrised by two structure functions $g_1$ and $g_2$, where $g_2$ is negligible and $g_1$ is given by: 
\begin{eqnarray}\label{g1-lo}
\nonumber g_1^{p,n}(x,Q^2)=\frac{1}{2} \sum_{q} e^2_q \left[
\Delta q^{p,n}(x,Q^2) + \Delta \bar{q}^{p,n}(x,Q^2)\right]
\, .
\end{eqnarray}
Here $\langle e^2\rangle=\sum_q e^2_q/N_q$ is the average squared charge of
all involved quark flavours, and $\Delta{q}(x,Q^2) = q_\sparq(x,Q^2) -
q_\santq(x,Q^2)$ is the quark helicity distribution for
massless
quarks of flavour $q$ in a longitudinally polarised
nucleon in the ``infinite momentum frame''.
\par
The structure function $g_1$ is related directly 
to the cross section  difference:
$\sigma_{LL} \equiv  \frac{1}{2}(\sigma^\sant - \sigma^\spar)/2 \,$,
where longitudinally ($L$) polarised leptons ($\rightarrow$) scatter on
longitudinally  ($L$) polarised nuclear targets with polarisation direction
either parallel  or anti-parallel ($\spar$, $\sant$) to the spin
direction of the beam.
The relationship to spin structure functions is:
\begin{equation} \label{eq:sigLL}
\nonumber
\frac{\detwo\sigma_{LL}(x,Q^2)}{\de x \de Q^2} =
\frac{8\pi\alpha^2y}{Q^4} 
\times \left[
\left( 1-\frac{y}{2} -\frac{y^2}{4}\gamma^2\right)\, g_1(x,Q^2) 
- \frac{y}{2}\gamma^2 \, g_2(x,Q^2) \right]\,,
\end{equation}
where $\gamma^2=Q^2/\nu^2$, with $\nu$ the energy of the virtual photon in the target rest frame.
\par
Measurements of $g_1$ for the proton, deuteron and neutron are shown in figure~\ref{fig:g1}(left) (from~\cite{Airapetian:2007mh}).
They can be used to extract the contribution of sea and valence quarks to the proton spin. Within some theoretical assumptions, this contribution is found to be $\Delta \Sigma_{(Q^2=5~{\rm GeV^2})}\simeq \int dx g_1(x,Q^2) \simeq 0.33$, which leaves a significant fraction for the gluon contribution to the proton spin. 
\par
The gluon contribution to the proton spin can in principle be extracted, similarly to the unpolarised case, from the scaling violations of the polarised structure functions. Unfortunately, in the phase space accessed by the fixed target experiments - relatively large $x$ and low $Q^2$ - scaling violations are expected to be small and the precision of the data limit this approach. An alternative method is based on the identification of hadrons in the final state with significant transverse momentum, enriching the selected sample in events initiated by direct gluon-photon interactions and  thereby enhancing the sensitivity of the analysis to the gluon polarisation $\Delta g/g$. A summary of the various analyses leading to estimations of the gluon contributions $\Delta g/g$ is presented in figure~\ref{fig:g1} (center, from~\cite{Alekseev:2008cz}). The precision of these extractions are still limited.
\par
The spin of the proton is also investigated in polarised $pp$ collisions at RHIC~\cite{Ellinghaus:2008eb}. Recent measurements of the cross section spin asymmetries for processes with inclusive jets in the final state open the way to new constraints, in particular for the gluon polarisation. The correlation between the observed jets allows for an approximate reconstruction of the kinematics. The measurement of the spin asymmetry as a function of the jet transverse momentum allows the investigation of the integrated gluon contribution to the proton spin $\Delta G(Q^2)=\int dx \Delta g(x,Q^2)$.
\par
Global fits have been performed recently with the aim of determining the polarised PDFs~\cite{deFlorian:2008mr} using  measurements from polarised inclusive and semi-inclusive DIS and from polarised proton-proton collisions. The analysis shows that the RHIC data have a good sensitivity to $\Delta g$. The measurement of the double helicity asymmetry $A_{LL}=(\sigma_{++}-\sigma_{+-})/\sigma_{++}-\sigma_{+-})$ is compared with the fit in figure~\ref{fig:g1}(right). The fit indicates a positive $\Delta g$ in a restricted region of $x$. The global fits are expected to also incorporate  the analysis of  final states with photons and $W$ bosons. These analyses have a significant potential to enhance the sensitivity of polarised $pp$ collision data to the gluon polarisation. In addition, the data collected from polarised $pp$ collisions at higher energy $\sqrt{s_{pp}}=500$~GeV and lower energy $\sqrt{s_{pp}}=64$~GeV~\cite{Adare:2008qb}  will allow to explore a larger $x$ domain, in particular at low $x$ where $\Delta g$ is unconstrained at present. The new measurements together with the expected refinements of the global fit procedures are expected to significantly improve  the knowledge of polarised PDFs in the future.


\section{Outlook}
 The research of  nucleon structure is the scene of fast progress. Precise data from HERA constrain the parton distribution functions to an unprecedented level and provide solid support for predictions at the  LHC in most of the phase space. The picture is complemented by Tevatron data at large $x$. The global fits exploit - and also prove - the universality of the parton distribution functions, combining the most precise results, the yield of which will increase in the years to come. Knowledge of the proton structure is profoundly enriched  by spin studies in polarised collisions. Future proposed projects at higher energies like LHeC~\cite{Dainton:2006wd} and EIC~\cite{Deshpande:2005wd}, have the potential sharpen significantly the precision of nucleon structure investigations and to bring this fundamental research area into new paradigms.

\end{document}